\documentclass{article}
\usepackage{spconf,amsmath,graphicx,hyperref}
\usepackage{cite}
\usepackage{amsmath,amssymb,amsfonts}
\usepackage{algorithmic}
\usepackage{graphicx}
\usepackage{multirow}
\usepackage{textcomp}
\usepackage{xcolor}
\usepackage{graphicx}
\usepackage{subcaption}
\usepackage{makecell}
\usepackage{hyperref}
\usepackage{booktabs}

\usepackage{tabularx} 
\newcolumntype{Y}{>{\centering\arraybackslash}X} 

\definecolor{DodgerBlue3}{RGB}{24, 116, 205} 
\definecolor{DarkOliveGreen4}{RGB}{110, 139, 61}  
\def\BibTeX{{\rm B\kern-.05em{\sc i\kern-.025em b}\kern-.08em
    T\kern-.1667em\lower.7ex\hbox{E}\kern-.125emX}}

\title{Multi-Channel Speech Enhancement for Cocktail Party Speech \\ Emotion Recognition}

\name{
\begin{tabular}{c}
Youjun Chen$^{\ast,1}$,
Guinan Li$^{\ast,1}$,
Mengzhe Geng$^2$, 
Xurong Xie$^{\dagger,3}$,
Shujie Hu$^1$, 
Huimeng Wang$^1$, 
Haoning Xu$^1$, \\
Chengxi Deng$^1$, 
Jiajun Deng$^1$,
Zhaoqing Li$^1$,
Mingyu Cui$^1$,
Xunying Liu$^{\dagger,1}$
\end{tabular}\thanks{ $^{\ast}$ Equal contribution: \{yjchen,gnli\}@se.cuhk.edu.hk}\thanks{$^{\dagger}$ Corresponding author: xyliu@se.cuhk.edu.hk;xurong@iscas.ac.cn}
}

\address{$^1$ The Chinese University of Hong Kong, Hong Kong SAR, China \\
$^2$ National Research Council Canada, Canada \\
$^3$ Institute of Software, Chinese Academy of Sciences, China}
\begin{document}
\ninept
\maketitle
\begin{abstract}
This paper highlights the critical importance of multi-channel speech enhancement (MCSE) for speech emotion recognition (ER) in cocktail party scenarios. A multi-channel speech dereverberation and separation front-end integrating DNN-WPE and mask-based MVDR is used to extract the target speaker's speech from the mixture speech, before being fed into the downstream ER back-end using HuBERT- and ViT-based speech and visual features. Experiments on mixture speech constructed using the IEMOCAP and MSP-FACE datasets suggest the MCSE output consistently outperforms domain fine-tuned single-channel speech representations produced by: {\bf a)} Conformer-based metric GANs; and {\bf b)} WavLM SSL features with optional SE-ER dual task fine-tuning. Statistically significant increases in weighted, unweighted accuracy and F1 measures by up to 9.5\%, 8.5\% and 9.1\% absolute (17.1\%, 14.7\% and 16.0\% relative) are obtained over the above single-channel baselines. 
The generalization of IEMOCAP trained MCSE front-ends are also shown when being zero-shot applied to out-of-domain MSP-FACE data\footnote{Demos are available at \href{https://SEUJames23.github.io/MCSE-ER/}{https://SEUJames23.github.io/MCSE-ER/}}.
\end{abstract}

\begin{keywords}
Emotion Recognition, Multi-channel Speech Enhancement, SSL Representations, Cocktail Party, Zero-shot
\end{keywords}
\vspace{-0.2cm}
\section{Introduction}
\vspace{-0.1cm}
\label{sec:intro}
Despite the rapid progress of speech emotion recognition (ER) in the past few decades, accurate ER of cocktail party mixture speech \cite{qian2018past} in far-field complex acoustic scenarios, e.g.,  driver anger expression detection in a noisy vehicle environment \cite{10913208}, patient emotion detection in noisy hospital \cite{tariq2019speech} etc., remains a highly challenging task to date \cite{conf/interspeech/LatifRKJS20}. 
Its difficulty can be attributed to multiple sources of interference including overlapping speakers, background noise and room reverberation. These lead to a large mismatch between the resulting mixture speech and clean signals.


Efforts to improve the performance of ER under noisy environments have been ongoing over the years. 
These include, but not limited to: \textbf{1)} data augmentation strategies like noise and speed perturbation \cite{conf/icassp/XuZCZ21,conf/icassp/DangVNW23}; \textbf{2)} domain fine-tuned SSL representations \cite{conf/icassp/NainiKRRB24,Tzeng2025}; \textbf{3)} using single-channel SE modules before ER back-end \cite{journals/taslp/LeemFOGB24,CHEN2024110169}. 
However, these prior studies suffer from several limitations: 
\textbf{a)} Lack of multi-channel speech enhancement (MCSE) for ER in cocktail party scenarios. Instead, the vast majority of current ER systems use single-channel noisy speech input only \cite{conf/icassp/XuZCZ21,conf/icassp/DangVNW23,conf/icassp/NainiKRRB24,Tzeng2025,journals/taslp/LeemFOGB24,CHEN2024110169}. The superiority of multi-channel beamforming over single-channel SE approaches has been widely shown in a wide range of speech processing tasks such as speech separation \cite{conf/icassp/Zhang00ZC021}, dereverberation \cite{conf/interspeech/YuS12}, recognition \cite{journals/taslp/LiDGJWHCML23,li2024joint} and diarization \cite{shao2024multichannelmultispeakerasrusing}. \textbf{b)} Lack of a complete multi-channel speech dereverberation and separation based SE front-end for downstream ER tasks. While a few recent researches \cite{conf/interspeech/GragedaAMBY23,conf/interspeech/GragedaAMBY24} provided some first insights into the application of multi-channel speech separation for distant ER, the impact of reverberation \cite{8645989} on ER was not considered. 
\textbf{c)} 
The efficacy of SE front-end was predominantly evaluated on audio-only ER systems \cite{journals/taslp/LeemFOGB24,CHEN2024110169} alone, while there is a lack of holistic evaluation of SE front-end on both audio-only and audio-visual ER systems.

In recent years, there has been a trend of conventional speech dereverberation approaches evolving into their current DNN-based variants, including: 
\textbf{a)} DNN-based weighted prediction error (DNN-WPE) \cite{conf/icassp/HeymannDHKN19} methods and \textbf{b)} spectral masking \cite{conf/icassp/FuLLLLJX22} and spectral mapping \cite{conf/icassp/WangW20} approaches.
Similarly, DNN-based microphone array beamforming techniques represented by \textbf{1)} neural time-frequency (TF) masking approaches \cite{conf/interspeech/BahmaninezhadWG19}; \textbf{2)} neural Filter and Sum methods \cite{7859320,conf/icassp/XiaoWELHSCZMY16}; and \textbf{3)} mask-based minimum variance distortionless response (MVDR) approaches \cite{conf/icassp/YoshiokaECA18} have been widely adopted for speech separation.   Furthermore, a pipelined MCSE front-end, integrating DNN-WPE based speech dereverberation and mask-based MVDR speech separation in our previous work \cite{journals/taslp/LiDGJWHCML23}, offers the best overall performance in both SE and ASR tasks.

To this end, this paper highlights the critical importance of MCSE for ER in cocktail party scenarios. A multi-channel DNN-WPE based speech dereverberation and mask-based MVDR speech separation front-end is used to extract the target speaker’s speech from the mixture speech, before being fed into the purpose-built downstream ER back-end using HuBERT \cite{journals/corr/abs-2106-07447} and ViT \cite{conf/iclr/DosovitskiyB0WZ21} based speech and visual features. Experiments on mixture speech constructed using the IEMOCAP \cite{Busso2008} and MSP-FACE \cite{Vidal_2020} datasets suggest the MCSE output consistently outperforms domain fine-tuned single-channel speech representations produced by: \textbf{a)} Conformer-based metric GANs; and \textbf{b)} WavLM SSL features with optional SE-ER dual-task fine-tuning.
Statistically significant increases in weighted, unweighted accuracy and F1 measures by up to \textbf{9.5\%, 8.5\%} and
\textbf{9.1\% absolute} (\textbf{17.1\%, 14.7\%} and \textbf{16.0\% relative}) for audio-only ER back-ends and \textbf{3.4\%, 3.9\%} and
\textbf{3.9\% absolute} (\textbf{4.5\%, 5.2\%} and \textbf{5.2\% relative}) for audio-visual ER back-ends are obtained over the corresponding single-channel based ER baselines on the IEMOCAP dataset.
Consistent improvements of speech quality on SRMR, PESQ and STOI scores are obtained.
The generalization of IEMOCAP trained MCSE front-ends are also shown when being zero-shot applied to out-of-domain MSP-FACE data.


\begin{figure*}[htbp]
\vspace{-0.6cm}
\centering
\setlength{\abovecaptionskip}{0pt plus 1pt minus 3pt}
\includegraphics[scale=0.40]{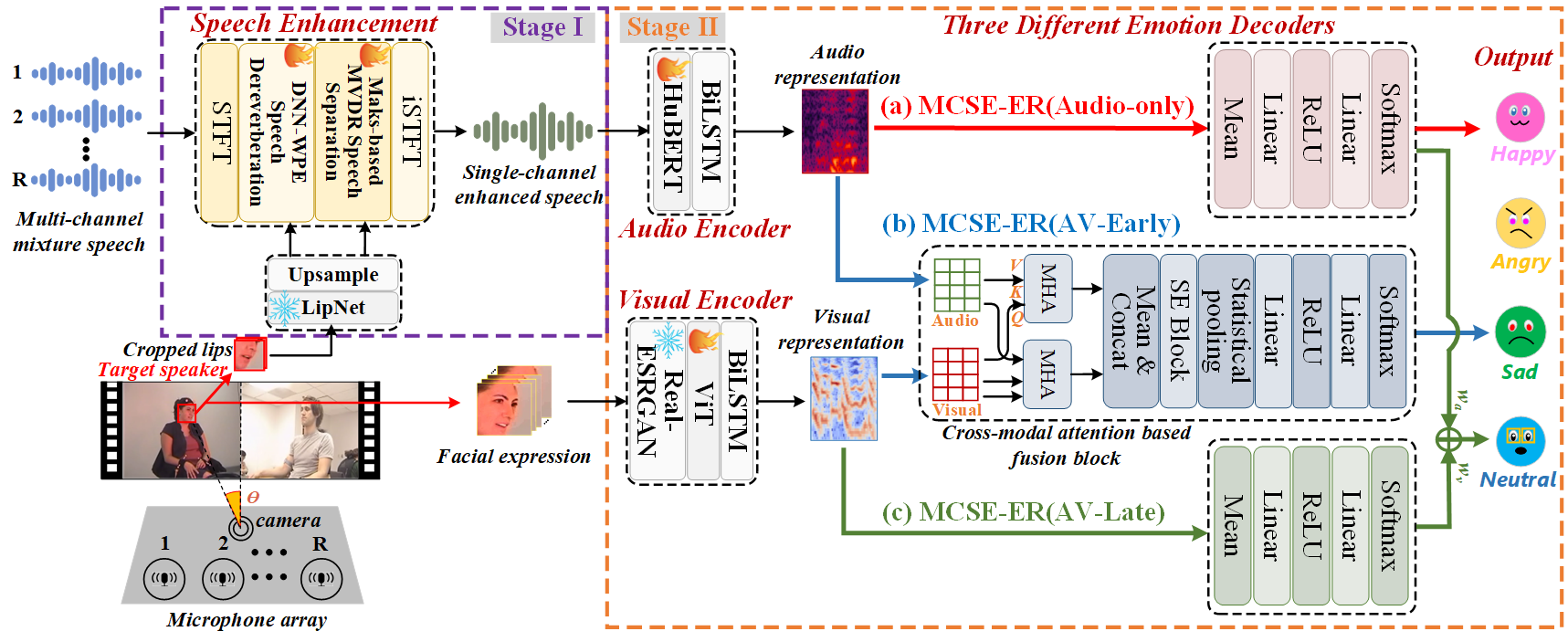}
\caption{Illustration of multi-channel speech enhancement (MCSE) and emotion recognition (ER) system. The MCSE front-end integrates DNN-WPE based speech dereverberation and mask-based MVDR speech separation components. Three different emotion recognition decoders are designed for audio-only and audio-visual ER systems, including: \textcolor{red}{\textbf{(a)} audio-only ER system (MCSE-ER (Audio-only))}, \textcolor{DodgerBlue3}{\textbf{(b)} early-fusion audio-visual ER system (MCSE-ER (AV-Early))} using cross-modal attention-based fusion block and \textcolor{DarkOliveGreen4}{\textbf{(c)} late-fusion audio-visual ER system (MCSE-ER (AV-Late))} using the weights of audio \textcolor{DarkOliveGreen4}{($w_{a}$)} and visual \textcolor{DarkOliveGreen4}{($w_{v}$)} to calculate the final classified probability.
}
\label{fig1}
\vspace{-0.6cm}
\end{figure*}

The main contributions of our work are summarized below:  

\textbf{1)} This paper highlights the critical importance of the MCSE front-end for the downstream ER back-end in cocktail party scenarios. 
In contrast, the vast majority of prior researches \cite{conf/icassp/XuZCZ21,conf/icassp/DangVNW23,conf/icassp/NainiKRRB24,Tzeng2025,journals/taslp/LeemFOGB24,CHEN2024110169} primarily focused on single-channel SE approaches for ER, with only a few exceptions \cite{conf/interspeech/GragedaAMBY23,conf/interspeech/GragedaAMBY24} in recent research on MCSE for ER task.

\textbf{2)} This paper completely analyses the efficacy of the MCSE front-end for both the audio-only and audio-visual ER back-ends. In contrast, previous studies only investigated the effect of the SE front-ends on the audio-only ER back-ends \cite{conf/icassp/XuZCZ21,conf/icassp/DangVNW23,conf/icassp/NainiKRRB24,Tzeng2025,journals/taslp/LeemFOGB24,CHEN2024110169,conf/interspeech/GragedaAMBY23,conf/interspeech/GragedaAMBY24}.

\textbf{3)} Detailed ablation studies demonstrate that our MCSE front-end, integrating both the DNN-WPE based speech dereverberation and mask-based MVDR speech separation components, can consistently improve the performance of the SE front-end and ER back-end. In contrast, limited related studies \cite{conf/interspeech/GragedaAMBY23,conf/interspeech/GragedaAMBY24} only explored the efficacy of multi-channel speech separation for cocktail party speech ER, but overlooked the importance of both components.

\textbf{4)} This paper explores zero-shot performance of MSCE front-end for ER in cocktail party scenarios. In contrast, the vast majority of prior researches \cite{conf/icassp/XuZCZ21,conf/icassp/DangVNW23,conf/icassp/NainiKRRB24,Tzeng2025,journals/taslp/LeemFOGB24,CHEN2024110169,conf/interspeech/GragedaAMBY23,conf/interspeech/GragedaAMBY24} lacks studies on this aspect.

\vspace{-0.2cm}
\section{Single-Channel Speech Enhancement}
\vspace{-0.1cm}
\label{SCSE}

The following two representative single-channel SE approaches for noisy ER are used as baselines in this paper:

\noindent \textbf{1) The Conformer-based Metric GAN} (CMGAN) approach \cite{cao2022cmgan} for ER is based on either \textbf{a)} the pre-trained CMGAN model (Sys. 3 in Tab. \ref{table1}) referring to the prior research \cite{CHEN2024110169}, or \textbf{b)} the domain mixture speech fine-tuned CMGAN (Sys. 4 in Tab. \ref{table1}).

\noindent \textbf{2) WavLM SSL features} are extracted from the domain mixture speech fine-tuned WavLM\footnote{WavLM is pre-trained on the simulated cocktail party speech data, enabling the model with the ability to perform speech denoising.} \cite{chen2022wavlm} model for ER (Sys. 1 in Tab. \ref{table1}) or further fine-tuned by the SE-ER dual task (Sys. 2 in Tab. \ref{table1}) as proposed in the prior study \cite{Tzeng2025}.

\vspace{-0.3cm}
\section{Multi-Channel Speech Enhancement}\label{mcse}
\vspace{-0.2cm}
A pipelined multi-channel speech enhancement (MCSE) front-end integrating DNN-WPE based speech dereverberation followed by mask-based MVDR speech separation \cite{journals/taslp/LiDGJWHCML23} (Fig. \ref{fig1}, top left, in light yellow) is introduced before being connected to the emotion recognition back-end of Section \ref{sec:ER_back_end}.

\subsection{DNN-WPE for Speech Dereverberation}\label{DWSD}
\vspace{-0.2cm} 
Let $R$ represents the channel number of a microphone array, and let $t$ and $f$ respectively denote the indices of time and frequency bins. 
The reverberant and overlapped multi-channel speech spectrum vector $\mathbf{x}(t, f) \in \mathbb{C}^{R} $ is first processed by the speech dereverberation component.
The dereverberated multi-channel speech spectrum vector $\hat{\mathbf{d}}(t, f) \in \mathbb{C}^{R} $ is obtained by applying the WPE filter $\mathbf{W}_{\text{\tiny{WPE}}}(f) \in \mathbb{C}^{LR \times R}$ to the time-delayed reverberant speech spectrum vector $\tilde{\mathbf{x}}(t\!-\!D, f)$ as follows:
\begin{align} 
\label{equation:dnn_wpe_filtering}
\hat{\mathbf{d}}(t, f) &= \mathbf{x}(t, f) - \mathbf{W}_{\text{\tiny{WPE}}}(f)^H \tilde{\mathbf{x}}(t\!-\!D, f),
\end{align}
where $\tilde{\mathbf{x}}(t\!-\!D,\!f) \!=\! [\mathbf{x}(t\!-\!D, \!f)^{T}, \ldots, \mathbf{x}(t\!-\!D\!-\!L\!+\!1, f)^{T} ]^{T}$, while $D$ and $L$ denote the prediction delay parameter and the number of filter taps, respectively.

The WPE filter $\mathbf{W}_{\text{\tiny{WPE}}}(f)$ is derived using the DNN-WPE approach \cite{conf/interspeech/KinoshitaDKMN17}, where the filtered signal power $\lambda(t,f)$ is estimated using DNN-predicted TF complex mask $M_{\text{\tiny{WPE}}}(t,f) \in \mathbb{C}$ as follows:
\begin{small}
\begin{align}  
\label{equation:dnn_wpe_filter}
 \mathbf{W}_{\text{\tiny{WPE}}}(f) = {\left(\sum_{t}\frac{\tilde{\mathbf{x}}(t\!-\!D, f) \tilde{\mathbf{x}}(t\!-\!D, f)^{H}}{\lambda(t,f)}\right)^{-1}} \nonumber \\ \left(\sum_{t}\frac{\tilde{\mathbf{x}}(t\!-\!D, f) \mathbf{x}(t, f)^{H}}{\lambda(t,f)}\right),
\end{align}
\end{small}

\begin{equation} 
\label{equation:dnn_wpe_psd}
 \lambda(t,f) = \frac{1}{R} \Vert M_{\text{\tiny{WPE}}}(t,f) \mathbf{x}(t,f) \Vert_2^2,
\end{equation}
where $\Vert\cdot\Vert_2$ denotes the Euclidean norm. 
The above alternating estimation procedure is iterated until convergence. 

\vspace{-0.3cm}
\subsection{Mask-based MVDR for Speech Separation}\label{MMSS}
\vspace{-0.1cm}
In MVDR beamforming \cite{conf/icassp/YoshiokaECA18}, a linear filter $\mathbf{W}_{\text{\tiny{MVDR}}}(f) \in \mathbb{C}^{R}$ is applied to the prior dereverberated multi-channel output $\hat{\mathbf{d}}(t, f)$ to produce the filtered single-channel enhanced speech spectrum as:
\begin{align} 
\hat{S}(t, f) = \mathbf{W}_{\text{\tiny{MVDR}}}(f)^H \hat{\mathbf{d}}(t, f),\label{equation:linear_filtering_1}
\end{align}
where $(\cdot)^{H}$ denotes the conjugate transpose operator. 

By minimizing the residual noise output while imposing a distortionless constraint on the target speech, 
the MVDR beamforming filter is estimated as:
\begin{align}
{\small \mathbf{W}_{\text{\tiny{MVDR}}}(f)\!=\frac{\boldsymbol{\Phi}_n(f)^{-1} \boldsymbol{\Phi}_x(f)} {\operatorname{tr}\left(\boldsymbol{\Phi}_n(f)^{-1} \boldsymbol{\Phi}_x(f)\right)} \mathbf{u}_{r}},\label{equation:mvdr_filter_2}
\end{align}
where $\mathbf{u}_r=[0,\ldots, 1, \ldots, 0]^T \in \mathbb{R}^{R}$ is a one-hot reference vector with the $r$-th component set to one. 
$\operatorname{tr}(\cdot)$ denotes the trace operator.
Without loss of generality, we select the first channel as the reference in this paper.
The target speaker power spectral density (PSD) matrix $\boldsymbol{\Phi}_x(f)$  and the noise-specific PSD matrix $\boldsymbol{\Phi}_n(f)^{-1}$ are computed using DNN-predicted complex TF masks \cite{journals/taslp/LiDGJWHCML23}. 


The MCSE front-end integrating DNN-WPE based speech dereverberation and mask-based MVDR speech separation, can optionally incorporate lip visual features as proposed in prior research \cite{journals/taslp/LiDGJWHCML23}.

\vspace{-0.4cm}
\section{Emotion Recognition Back-end}
\vspace{-0.2cm}
\label{sec:ER_back_end}
\subsection{Audio Encoder}
\vspace{-0.2cm}
The single-channel enhanced speech produced by the  MCSE front-end is fed into the audio encoder (Fig. \ref{fig1}, top middle, in light gray) to extract and downsample the audio representation, where the audio encoder comprises a pre-trained SSL model HuBERT \cite{journals/corr/abs-2106-07447} encoder and a BiLSTM layer.
Specifically, the audio representation is obtained from the output of the final transformer layer of HuBERT referring to \cite{conf/icassp/NainiKRRB24}.
The subsequent BiLSTM layer is employed to reduce the dimensionality of the audio representation while preserving temporal information.

\vspace{-0.4cm}
\subsection{Visual Encoder}
\vspace{-0.2cm}
Facial expressions of the target speaker are processed by the visual encoder (Fig. \ref{fig1}, bottom middle, in light gray), which comprises Real-ESRGAN \cite{conf/eccv/WangYWGLDQL18}, the pre-trained ViT \cite{conf/iclr/DosovitskiyB0WZ21} encoder, and BiLSTM modules.
Similar to the audio encoder, visual representations are extracted from the output of the ViT's last transformer layer and then downsampled to match the dimensionality of the downsampled audio representation for further modality fusion.

\begin{table*}
 \vspace{-1.5cm}
\caption{
Performance comparison between multi-channel speech enhancement (MCSE) front-end and single-channel speech enhancement baselines for audio-only (Sys. 5 vs. Sys.1-4) and audio-visual (Sys. 10 vs. Sys.6-9) emotion recognition (ER) on multi-channel IEMOCAP and MSP-FACE datasets. When these systems are evaluated on MSP-FACE, the IEMOCAP trained SE front-end is directly used to produce enhanced speech for ER back-end fine-tuning, marked as ``zero-shot SE" in the last column of this table.
``$\ast$'', ``$\dagger$", ``$\ddagger$'' and ``$\circ$'' represent statistically significant (Paired Single-tailed T-test \cite{champoux2022first}, $p$=0.05) accuracy improvements over Sys.2, Sys.4, Sys.7 and Sys.9, respectively.}
\centering
\setlength{\tabcolsep}{1pt} 
\fontsize{8.3}{10}\selectfont 
\begin{tabularx}{\textwidth}{c|Y|c|c|c|c|ccc|ccc|ccc}
    \toprule[1.2pt]
    \multirowcell{5}{ID} & \multirowcell{5}{System} & \multicolumn{4}{c|}{Experimental Setup} & \multicolumn{3}{c|}{SE Front-end} & \multicolumn{6}{c}{ER Back-end} \\
    \cline{3-15}
    & & \multirowcell{3}{\scriptsize{SE} \\ \scriptsize{Front-end}} & \multicolumn{3}{c|}{\multirowcell{3}{\scriptsize{ER Back-end}}} & \multicolumn{3}{c|}{\multirowcell{3}{\scriptsize{Trained \& Evaluated} \\ \scriptsize{on IEMOCAP}}} & \multicolumn{3}{c|}{\multirowcell{3}{\scriptsize{Fine-tuned \& Evaluated} \\ \scriptsize{on IEMOCAP}}} & \multicolumn{3}{c}{\multirowcell{3}{\scriptsize{Fine-tuned \& Evaluated} \\ \scriptsize{on MSP-FACE}\\ \scriptsize{(Via zero-shot SE)}}} \\
    & & &\multicolumn{3}{c|}{}&&&&&&&&& \\
    & & &\multicolumn{3}{c|}{}&&&&&&&&& \\
    \cline{3-15}
     & & \scriptsize{\#Chn.} & \scriptsize{Audio Enc.} & \scriptsize{Visual Enc.} & \scriptsize{AV Fusion} & \scriptsize{SRMR $\uparrow$} & \scriptsize{PESQ $\uparrow$} & \scriptsize{STOI $\uparrow$} & \scriptsize{WA\% $\uparrow$} & \scriptsize{UA\% $\uparrow$} & \scriptsize{F1\% $\uparrow$} & \scriptsize{WA\% $\uparrow$} & \scriptsize{UA\% $\uparrow$ } & \scriptsize{F1\% $\uparrow$} \\
    \midrule[1pt]
    \multicolumn{6}{c|}{\scriptsize{The Raw First Channel of Mixture Speech}} & 3.16 & 1.16 & 0.48 & \multicolumn{6}{c}{/}\\
    \hline
    1& \scriptsize{WavLM \cite{chen2022wavlm} + ER fine-tuning} & \multirowcell{4}{\scriptsize{Single}} & \multirowcell{2}{\scriptsize{WavLM}} & \multirowcell{5}{/} & \multirowcell{5}{/} & \multicolumn{3}{c|}{/} & 54.3 & 55.6 & 55.1 & 38.4 & 32.1 & 31.9 \\
    2 &  \scriptsize{WavLM + SE-ER fine-tuning \cite{Tzeng2025}} & & &  &  & 2.91 & 1.18 & 0.51 & 55.7 & 57.7 & 56.8 & 40.6 & 33.8 & 33.5 \\
    \cline{2-2}
    \cline{4-4}
    \cline{7-15}
    3&  \scriptsize{CMGAN \cite{cao2022cmgan} + HuBERT} &  & \multirowcell{3}{\scriptsize{HuBERT}} & & & 3.65 & 1.27 & 0.60 & 56.5 & 58.3 & 57.7 & 38.6 & 32.5 & 32.2 \\
    4& \scriptsize{Fine-tuned CMGAN + HuBERT} & & &  &  & 3.88 & 1.42 & 0.64 & 57.1 & 58.0 & 57.6 & 41.2 & 34.6 & 33.8 \\
    \cline{2-3}
    \cline{7-15}
    5 &  \scriptsize{\textbf{MCSE + HuBERT(ours)}} &  \scriptsize{Multi} & & &  & \textbf{6.69} & \textbf{2.82} & \textbf{0.76} & \textbf{65.2} & \textbf{66.2} & \textbf{65.9}$^{\ast}$$^{\dagger}$ & \textbf{43.9} & \textbf{37.4} & \textbf{36.3}$^{\ast}$$^{\dagger}$ \\
    \hline
    \hline
    6& \scriptsize{WavLM \cite{chen2022wavlm} + ER fine-tuning} & \multirowcell{4}{\scriptsize{Single}} & \multirowcell{2}{\scriptsize{WavLM}} & \multirowcell{5}{\scriptsize{ViT}} & \multirowcell{5}{\scriptsize{Early-}\\\scriptsize{Fusion}} & \multicolumn{3}{c|}{/} & 73.5 & 74.8 & 74.4 & 62.7 & 56.9 & 58.0 \\
    7& \scriptsize{WavLM + SE-ER fine-tuning \cite{Tzeng2025}} &  & &  & &2.91 & 1.18 & 0.51 & 74.9 & 75.6 & 75.3 & 64.3 & 58.5 & 59.3 \\
    \cline{2-2}
    \cline{4-4}
    \cline{7-15}
    8 & \scriptsize{ CMGAN \cite{cao2022cmgan} + HuBERT} &  & \multirowcell{3}{\scriptsize{HuBERT}} &  & &3.65 & 1.27 & 0.60 & 75.2 & 75.9 & 75.7 & 63.8 & 57.7 & 58.5 \\
    9& \scriptsize{Fine-tuned CMGAN + HuBERT} &  & &  & &3.88 & 1.42 & 0.64 & 75.5 & 76.1 & 75.9 & 65.1 & 59.2 & 60.3 \\
    \cline{2-3}
    \cline{7-15}
    10& \scriptsize{\textbf{MCSE + HuBERT(ours)}} & \scriptsize{Multi} & &  & &\textbf{6.69} & \textbf{2.82} & \textbf{0.76} & \textbf{78.3} & \textbf{79.5} & \textbf{79.2}$^{\ddagger}$$^{\circ}$ & \textbf{67.4}& \textbf{61.4} & \textbf{62.3}$^{\ddagger}$$^{\circ}$ \\
    \midrule[1.2pt]
\end{tabularx}
\label{table1}
 \vspace{-0.7cm}
\end{table*}

\vspace{-0.4cm}
\subsection{Three Different Emotion Decoders}
\vspace{-0.2cm}
To explore the efficacy of the MCSE front-end for both audio-only and more competitive audio-visual ER back-ends, three different purpose-built emotion decoders are designed. These include:
\textbf{a)} In the audio-only system (MCSE-ER (Audio-only), Fig. \ref{fig1}, top right, in light red), the audio representation is directly used to predict emotion labels;
\textbf{b)} In the early-fusion audio-visual system (MCSE-ER(AV-Early), Fig. \ref{fig1}, middle right, in light blue), the extracted audio and visual representations are fed into the cross-modal attention-based fusion block\footnote{Cross-modal attention-based fusion block consists of two multi-head attention (MHA) layers with 6 heads, mean \& concatenation module, squeeze \& excitation (SE) block \cite{conf/asru/LiuJN23}, and a statistical pooling layer.} to generate joint audio-visual representations for emotion classification;
and \textbf{c)} In the late-fusion audio-visual system ((MCSE-ER(AV-Late), Fig. \ref{fig1}, bottom right, light green), emotion labels are predicted using a learnable weighted sum of probabilities separately predicted by audio and visual representations. 

\vspace{-0.3cm}
\section{Experiments}
\vspace{-0.3cm}

\subsection{Task Description}\label{dp}
\vspace{-0.1cm}
\textbf{IEMOCAP corpus \cite{Busso2008}} is a single-channel multi-modal multi-speaker dataset, and widely used in ER research.
To ensure consistency with prior studies \cite{journals/corr/abs-2406-07162}, we focus on a 4-way classification task by excluding utterances with labels outside the set \{$neutral$, $happy$, $sad$, $angry$, $excited$\}.
The $happy$ and $excited$ classes are further merged due to their expressive similarity. 
Finally, a clean dataset of 5,531 utterances with 4 different emotion labels is used for the following multi-channel mixture speech simulation.

\noindent \textbf{MSP-FACE corpus \cite{Vidal_2020}} is a real-world noisy single-channel emotional multi-modal corpus collected from video-sharing websites. Due to some videos being set to private, we ultimately obtain 732 utterances (95 videos) from training set and 434 utterances (60 videos ) from test set for multi-channel mixture speech simulation.

\noindent \textbf{Simulated multi-channel mixture speech:} The use of such data in the experiments is based on the following reasons: Firstly, prior researches on environment robust ER \cite{conf/icassp/XuZCZ21,conf/icassp/DangVNW23,conf/icassp/NainiKRRB24,Tzeng2025,journals/taslp/LeemFOGB24,CHEN2024110169,conf/interspeech/GragedaAMBY23,conf/interspeech/GragedaAMBY24} have widely used simulated noisy speech data due to lack of such data. Secondly, at present there is no publicly available real-recorded Cocktail Party speech dataset tailored for ER tasks. Thirdly, in a wider context, simulated noisy speech data is also used in the pre-training of state-of-the-art single-channel foundation models (e.g. WavLM \cite{chen2022wavlm}). Hence, we simulated the multi-channel mixture speech using the above two corpora following the prior simulation protocol \cite{journals/taslp/LiDGJWHCML23,li2024joint} for model training of the MCSE front-end and ER back-end. 
For each utterance, we uniformly sampled signal-to-noise ratio (SNR), signal-to-interference ratio (SIR) and reverberation time $T_{60}$ to construct cocktail party speech containing noise, overlapping speakers and reverberation\footnote{More details on the simulation process can be found in \cite{journals/taslp/LiDGJWHCML23}.}. 
The resulting simulated IEMOCAP-based mixture data contains 110.6k utterances, totalling 140 hours. For speaker-independent evaluation, we adopt the widely used 5-fold cross validation way \cite{journals/corr/abs-2406-07162}, splitting four sessions with eight speakers into training (67.7K utterances) and development (18.1K utterances) sets, while reserving one session with two speakers as the test set (24.8K utterances). 
The resulting simulated MSP-FACE-based mixture data contains a total of 14.6K utterances for training, and a total of 8.7K utterances for evaluation.

\noindent \textbf{Experimental design:}
The MCSE front-end and ER back-end are first trained and evaluated on multi-channel IEMOCAP data. To further assess the generalization of the MCSE front-end, the IEMOCAP trained MCSE front-ends have been zero-shot applied to out-of-domain MSP-FACE data.

\vspace{-0.4cm}
\subsection{Experimental Setup}\label{MC}
\vspace{-0.2cm}
\noindent \textbf{Model configuration:} Except for the pre-trained models Real-ESRGAN\footnote{https://github.com/xinntao/Real-ESRGAN}, HuBERT\footnote{https://huggingface.co/facebook/hubert-large-ls960-ft} and ViT\footnote{https://huggingface.co/dima806/facial\_emotions\_image\_detection}, the parameters of the other modules in the MCSE-ER systems are all trained from scratch. Specifically, MCSE front-ends adopt the same configuration as described in \cite{journals/taslp/LiDGJWHCML23} for DNN-WPE based speech dereverberation and masked-based MVDR speech separation.
The parameters of the HuBERT and ViT encoders are trainable, with the dimensionality of their last layers set to 1024 and 768, respectively.
Subsequently, their BiLSTM layers are utilized to reduce the dimensionalities of audio and visual representations to $2\times60$ while preserving temporal information.
During visual representation extraction, all-zero vectors are used instead when video-captured facial expressions are unavailable.

\noindent \textbf{Training strategy:}
We train the MCSE-ER systems in a pipelined two-stage way as the focus of the paper is to explore the importance of the MCSE front-end on the downstream ER task.
In the first stage (Purple dotted line, Fig. \ref{fig1}), the MCSE front-end is trained separately by maximizing the SISNR metric \cite{journals/taslp/LiDGJWHCML23}. 
In the second stage (Orange dotted line, Fig. \ref{fig1}), with the parameters of the MCSE front-end frozen, we fine-tune the entire system on the downstream ER task.

\noindent \textbf{Evaluation metrics:} To evaluate the performance of proposed methods and baselines, we adopt the commonly used metrics weighted accuracy (WA), unweighted accuracy (UA) and macro f1 score (F1) for ER tasks. In addition, we use speech-to-reverberation modulation energy ratio (SRMR) \cite{falk2010non}, perceptual evaluation of speech quality (PESQ) \cite{recommendation2001perceptual}, and short-time objective intelligibility (STOI) \cite{taal2011algorithm} to evaluate the quality of enhanced speech. Since we randomly sample interfering speech and other multi-channel simulation parameters in the process of generating simulated noisy speech data, the simulated mixture speech samples are independent of each other, which fulfills the condition of the Paired Single-tailed T-test \cite{champoux2022first}.

\noindent \textbf{Baseline systems:} All single-channel SE methods of Section \ref{SCSE} are used as baselines for ER. In addition, domain fine-tuned speech representations produced by these SE baselines are further fused with visual representations (Sys. 6-9 in Tab. \ref{table1})) based on the early-fusion method used in our \textbf{MCSE + Hubert} system (Sys. 10 in Tab. \ref{table1}).

\begin{table}
    \caption{Ablation studies on the importance of MCSE front-end components and audio-visual fusion approaches for ER on IEMOCAP data. ``$\ast$'', ``$\dagger$" and $\ddagger$ represent a statistically significant (Paired Single-tailed T-test, $p$=0.05) accuracy decrease over Sys.1, Sys.5 and Sys.9 with both speech dereverberation (dervb.) and separation (sep.) components in the MCSE front-ends, respectively.
    }
    \label{table2}
    \centering
    \resizebox{0.95\columnwidth}{!}{
    \begin{tabular}{c|c|ccc|ccc}
    \toprule[0.8pt]
    \multirowcell{2}{ID} & \multirowcell{2}{System} & \multicolumn{3}{c|}{SE Front-end} & \multicolumn{3}{c}{ER Back-end}\\
    \cline{3-8}
    &&\scriptsize{SRMR} & \scriptsize{PESQ}  & \scriptsize{STOI}  & \scriptsize{WA\%} & \scriptsize{UA\%} & \scriptsize{F1\%}\\
    \midrule[1pt]
    1& \textbf{MCSE-ER(Audio-only)} & \textbf{6.69} & \textbf{2.82} & \textbf{0.76} & \textbf{65.2} & \textbf{66.2} & \textbf{65.9} \\
    2& w/o dervb. & 5.52 & 2.56 & 0.70 & 63.2&63.9&64.0$^{\ast}$\\
    3& w/o sep. & 5.83 & 1.73 & 0.66 & 56.6&57.2&56.8$^{\ast}$\\
    4& w/o dervb. \& sep. & 3.16 & 1.16 & 0.48 & 52.5&54.2&53.2$^{\ast}$\\
    \hline
    5& \textbf{MCSE-ER(AV-Late)} & \textbf{6.69} & \textbf{2.82} & \textbf{0.76} & \textbf{75.9} &\textbf{78.1}&\textbf{77.5} \\
    6& w/o dervb. & 5.52 & 2.56 & 0.70 & 73.4&76.1&75.2$^{\dagger}$ \\
    7& w/o sep. & 5.83 & 1.73 & 0.66 & 71.2&74.7&73.4$^{\dagger}$ \\
    8& w/o dervb. \& sep. & 3.16 & 1.16 & 0.48 & 70.3&73.9&72.3$^{\dagger}$ \\
    \hline
    9& \textbf{MCSE-ER(AV-Early)} & \textbf{6.69} & \textbf{2.82} & \textbf{0.76} & \textbf{78.3} & \textbf{79.5} & \textbf{79.2} \\
    10& w/o dervb. & 5.52 & 2.56 & 0.70 & 75.4&77.7&76.6$^{\ddagger}$ \\
    11& w/o sep. & 5.83 & 1.73 & 0.66 & 72.8&75.3&74.5$^{\ddagger}$ \\
    12& w/o dervb. \& sep. & 3.16 & 1.16 & 0.48 & 71.5&74.1&73.4$^{\ddagger}$ \\
    \bottomrule[1.2pt]
    \end{tabular}
    }
\vspace{-0.6cm}
\end{table}
 \vspace{-0.4cm}
\subsection{Experimental Results}\label{results}
 \vspace{-0.1cm}
\noindent \textbf{Performance comparison between the MCSE front-end and single-channel SE front-end for audio-only ER} is shown in Tab. \ref{table1} (Sys. 1-5).
The MCSE output for audio-only ER (Sys. 5) consistently outperforms the domain fine-tuned single-channel speech representations produced by: a) Conformer-based metric GANs (Sys. 4); and b) WavLM SSL features with optional SE-ER dual task fine-tuning (Sys. 1,2). 
Statistically significant increases in weighted, unweighted accuracy and F1 measures by up to \textbf{9.5\%}, \textbf{8.5\%} and \textbf{9.1\% absolute} (\textbf{17.1\%}, \textbf{14.7\%} and \textbf{16.0\% relative}) are obtained (Sys. 5 vs. Sys. 2) on the IEMOCAPS data. 
Consistent performance improvements in speech enhancement metrics including SRMR, PESQ, and STOI are also obtained. Furthermore, the generalization of IEMOCAP trained MCSE front-ends are also shown when being zero-shot applied to out-of-domain MSP-FACE data with the statistically significant increases in weighted, unweighted accuracy and F1 measures by up to \textbf{5.5\%},  \textbf{5.3\%} and \textbf{4.4\%}
absolute (\textbf{14.3\%}, \textbf{16.5\%} and \textbf{13.8\%} relative) (Sys. 5 vs. Sys. 1).

 
\noindent \textbf{Performance comparison between the MCSE front-end and single-channel SE front-end for audio-visual ER} is shown in Tab. \ref{table1} (Sys. 6-10). 
The same trend can be found in the audio-visual ER task.
Specifically, the MCSE output for the audio-visual ER system (Sys.10) consistently outperforms the speech representations produced by the corresponding single-channel baselines (Sys. 6-9) on the IEMOCAP data with statistically significant increases in weighted, unweighted accuracy and F1 measures by up to \textbf{3.4\%}, \textbf{3.9\%} and \textbf{3.9\% absolute} (\textbf{4.5\%}, \textbf{5.2\%} and \textbf{5.2\% relative}) (sys. 10 vs. sys. 7).
This trend can also be found in MSP-FACE data, even using the zero-shot MCSE front-end.
Besides, the early-fusion method for audio-visual ER is used in these systems (Sys. 6-10) and its superiority is demonstrated in the following ablation studies.

\noindent \textbf{Ablation studies on the importance of MCSE front-end components and audio-visual fusion approaches} on IEMOCAP data are shown in Tab.\ref{table2}. \textbf{1)} The MCSE front-end integrating both the purpose-built DNN-WPE based speech dereverberation and mask-based MVDR speech separation components can consistently obtain the best performance, irrespective of in the audio-only or audio-visual ER systems (Sys. 1 vs. Sys. 2-4), (Sys. 5 vs. Sys. 6-8), (Sys. 9 vs. Sys. 10-12). Consistent performance improvements in speech enhancement front-end metrics scores were also obtained.
\textbf{2)} Based on the above best MCSE front-end setting, it can be observed that the early-fusion approach for audio-visual ER can consistently outperform the late-fusion approach (Sys. 9 vs. Sys. 5).

    


\vspace{-0.4cm}
\section{Conclusion}
\vspace{-0.3cm}
This paper highlights the critical importance of MCSE for ER in cocktail party scenarios. A multi-channel speech dereverberation and separation front-end integrating DNN-WPE and mask-based MVDR is used to extract the target speaker's speech from the mixture, before being fed into the downstream ER back-end using HuBERT- and ViT-based speech and visual features. Experiments on mixture speech constructed using the IEMOCAP dataset suggest the MCSE output consistently outperforms domain fine-tuned single-channel speech representations.
Statistically significant increases in weighted, unweighted accuracy and F1 measures by up to 9.5\%, 7.5\% and 8.4\% absolute (17.1\%, 13.0\% and 14.8\% relative) are obtained over the above single-channel baselines. The generalization of IEMOCAP trained MCSE front-ends are also shown when being zero-shot applied to out-of-domain MSP-FACE data. Future research will focus on developing multi-task systems that combine speech and emotion recognition in complex acoustic environments.

\vspace{-0.4cm}
\section{Acknowledgements}
\vspace{-0.3cm}
This research is supported by Hong Kong RGC GRF grant No. 14200220, 14200021, 14200324, Innovation Technology Fund
grant No. ITS/218/21, Research Project of Institute of Software, Chinese Academy of Sciences No. ISCAS-ZD-202401, ISCAS-JCMS-202306, and Youth Innovation Promotion Association CAS Grant No. 2023119.

\vspace{-0.3cm}
\bibliographystyle{IEEEbib}
\bibliography{strings,refs}

\end{document}